\documentstyle[12pt]{article}
\newcommand\dd{\mathrm{d}}

\title{On a general way of two-particle problem reduction}

\author{A.B. Arbuzov$^{1,2}$, T.V. Kopylova$^{2}$, A.B. Zhunisbayev$^{1}$}

\date{}

\begin{document}

\maketitle

\vskip 0.3cm

\begin{center}
{$^1$ \it Bogoliubov Laboratory for Theoretical Physics, JINR, Dubna, \\
141980 Russia\\
$^2$ \it Department of Higher Mathematics, University of Dubna, \\
141980 Dubna, Russia}
\end{center}

\begin{abstract}
The problem of the description of two interacting
particles is considered. It is shown that it can
be reduced to the description of one particle in 
an external static potential even in a relativistic case.
The method is based on the Maupertuis least action principle.
The proposed method is verified for known problems in two-particle problems
in classical mechanics, as well as in non-relativistic and relativistic 
quantum mechanics. 
\end{abstract}


\section{Introduction}

The problem of two--particle mutual interactions in physics is very old.
The traditional way to solve it is to consider it in
the {\em center-of-mass system} (c.m.s.) of reference~\cite{Newton}.
After a simple substitution of variables and excluding
the motion of the system as a whole, one obtains the
equation for an effective particle with a {\em reduced mass} in an
external potential of a fixed center.
It is well known how to do that in the classical mechanics (in the
two--body Kepler problem, for example) and in the non--relativistic
quantum mechanics in the two--particle Schr\"{o}dinger equation.
But in a relativistic case in the c.m.s. we do not have so clear
simplification of our problem. Even to write down a relativistic
Schr\"{o}dinger--like two--particle equation is a serious problem.
Of course there are many approaches to the relativistic
two--particle problem:
quasi--potential approach \cite{quasi} suggested by Logunov and
Tavkhelidze, the Bethe--Salpeter equation~\cite{BSeq},
the Droz--Vincent--Komar--Todorov equation~\cite{Droz,Komar,Tod} for scalar particles, 
the two-body Dirac equations~\cite{Saz,Crater:1987hm}
for spinor particles and many other. 

Below we propose a simple method which allows to reduce any two particle
problem to the description of an (effective) particle in a static central potential.
The idea is to exploit the duality between the scattering channel and the bound state
being the two branches of the same two-particle problem solution. 
The method works in the same way as in classical and 
quantum mechanics, as well as in the quantum field theory.
Our restrictions are natural: the particles
should be point--like (or can be effectively considered as
point--like) and their interaction should be local.

The paper is organized as follows.
In the second section of this paper we will consider the general
two--particle problem from the point of view of the least action
principle. From a class of possible ways of the elimination of
the degrees of freedom, corresponding to the motion of the
two--particle system as a whole, we will choose one.
As an illustration the classical Kepler problem will be considered.
In the third section we will imply our method to the following
problems: the two--particle Schr\"{o}dinger equation with the Coulomb
potential involved; the problem of electromagnetic interactions
between a spinor and a scalar relativistic charged particles; and
than --- the same problem for two spinor particles. The last section
of the paper contains conclusions and a discussion.

\section{The two--particle problem}

Hamilton's theorem proofs that the {\em two--body central force problem}
in classical mechanics 
can be reduced to the equivalent one--body problem, see e.g. Ref.~\cite{whitt}. 
The statement can be easily generalized for
non--relativistic quantum mechanics of two particles. But the reduction
of a relativistic two--particle problem still causes the appearance of
new approaches.

Let us suppose that we have a certain Lagrangian
for our two particles which includes an interaction. Usually one
constructs it in an inertial reference frame (RF) in order to avoid
additional "non--inertial" contributions. For an illustration we take now the classical
Kepler problem. The Lagrangian can be written as
\begin{eqnarray}
{\cal L} = \frac{m_{1} \dot{\vec{r}}_{1}\!^{2}}{2}
         + \frac{m_{2} \dot{\vec{r}}_{2}\!^{2}}{2} - U,\qquad
	 U= - \frac{m_1 m_2}{r},                       \label{lag}
\end{eqnarray}
where $m_1$ and $m_2$ are the masses of our particles,
$\vec{r}_1$ and $\vec{r}_2$ are their coordinates,
$r$ is the absolute value of the relative coordinate
$(r \equiv |\vec{r}_2 - \vec{r}_2|)$, and the dots over
$\vec{r}_1$ and $\vec{r}_2$ denote time derivatives.
We chose unity for the value of the coupling constant.

Then we can use the least action principle and write down the
equation for the action variation. We will use here the fact of
the energy conservation in our system and look for a static solution. 
So, we can use the original action definition due to Maupertuis, 
Euler and Lagrange instead of the Hamilton's one.
The so called {\em abbreviated action} of a conservative system
does not contain the time variable.
Using the variational principle for it, one can obtain
particle trajectories. The time dependence could be restored
after all. Note, that in the quantum mechanics in a stationary case,
one also often looks for observable quantities and forget about
the time dependence (in the Schr\"{o}dinger representation).

The abbreviated action for the Kepler problem reads
\begin{eqnarray}
    {\cal S} & = & \int \sqrt{2(E-U) [ m_1 ( \dd \vec{r}_1 )^2
                                     + m_2 ( \dd \vec{r}_2 )^2 ]}
\label{act} \\
(\dd \vec{r}_i)^2 & \equiv & (\dd x_i)^2 + (\dd y_i)^2 + (\dd z_i)^2,
\nonumber
\end{eqnarray}
where the integral has to be taken between two points on a trajectory.
$E$ is the energy of the two--particle system.

Our approach is based on the following statement: 
the problem of the action variation can be considered in
{\em any reference frame} even in a non--inertial one. That was shown
by Lagrange and follows just from the fact that the action is a scalar.
In other words, we have written the action in a certain RF
and after that we can take its variation in any reference frame.
And there is only one reason to make a concrete choice:
one should choose the RF being suitable for the given problem.

Following the tradition one can substitute the variables in Eq.~(2) as
\begin{eqnarray}
 \vec{r}_1,\ \vec{r}_2,\ \vec{p}_1,\ \vec{p}_2
\longrightarrow \vec{R},\   \vec{r},\ \vec{P},\ \vec{p},  \label{subs}
\end{eqnarray}
where
\begin{eqnarray}
&& \vec{R}=\frac{m_1 \vec{r}_1 + m_2 \vec{r}_2}{m_1 + m_2}\; , \qquad
\vec{r}=\vec{r}_2 - \vec{r}_1\; , \nonumber \\ \nonumber
&& \vec{P}=\vec{p}_1+\vec{p}_2\; , \qquad
\vec{p}=\frac{m_1 \vec{p}_2 - m_2 \vec{p}_1}{m_1 + m_2}\, .
\end{eqnarray}
The substitution for the particle momenta $\vec{p}_1$ and
$\vec{p}_2$ is shown for a completeness.
Further we choose the c.m.s. from all possible reference frames
and obtain the action in the form
\begin{eqnarray}
{\cal S} & = & \int \sqrt{2(E-U)  \mu } \; \dd \vec{r}, \qquad
\mu=\frac{m_1 m_2}{m_1 + m_2}\, .
\end{eqnarray}
This expression has the form of a one--particle abbreviated
action. The subsequent procedure is straight forward, we are not
going to repeat it in details. It is well known that the c.m.s.
is rather suitable for problems which have the same kinetic
part in their Lagrangians (for the two--particle Schr\"{o}dinger
equation, for instance). But in a relativistic case we have another
structure of the kinetic part. And the choice of the c.m.s. as
a reference frame does not reduce a relativistic two--particle
problem to a simple one--particle problem\footnote{Certainly, 
it reduces the number of the degrees of freedom but it does
not lead to the one--particle Dirac equation, for instance.}.

Let us consider another possible way to choose the RF.
The class of reference frames, where
\begin{eqnarray}
A \vec{p}_1 + B \vec{p}_2 =0,  \label{class}
\end{eqnarray}
has a simple physical interpretation. For example, choosing
$A=1$ and $B=0$ one obtains the rest frame of the first particle,
choosing $A=B=1$ one obtains the c.m.s.
The corresponding substitution for coordinates follows from the
conditions
\begin{eqnarray}
\vec{p}_1 \vec{r}_1  + \vec{p}_2 \vec{r}_2 = \vec{P} \vec{R}  +
\vec{p} \vec{r}\quad \mbox{and} \quad \vec{r}=\vec{r}_2-\vec{r}_1.
\end{eqnarray}
It is easy to show that all such choices of reference frames
are equivalent in a non--relativistic case: they all reduce the problem
to a one--particle one (the difference would be only in different
expressions for {\em reduced masses}), and the subsequent turn back
to an inertial (c.m.s.) system of reference is rather simple, it will be
discussed below. We will do now a concrete choice. We choose as
the working reference frame the rest frame of one of the particles,
of the first one to be definite.
There the two--particle action presented in eq. (\ref{act}) takes
the form
\begin{eqnarray}
{\cal S} =  \int \sqrt{2(E-U) m_2 } \; \dd \vec{r}_2, 
\label{effact}
\end{eqnarray}
which is just the form of the action for the motion of the second
particle in the potential created by the first one.
It is a certain degenerated case, the reduced mass now
coincides with the mass of the second particle.
One can see that the expression does not depend on the
relation between the masses. We would
obtain the same action as for the case $m_1 \gg m_2$ as well as
for the case $m_1 \ll m_2$. Solving the Euler--Lagrange equations
corresponding to the action we obtain a trajectory (a conic curve)
which depends actually on three parameters:
the energy $E_2$, the angular momentum $M_2$ and the mass $m_2$.
In the polar coordinates we have
\begin{equation}
\frac{M_2^2}{m_2 |\vec{r}_2|}
=1+\sqrt{1+\frac{2E_2 M_2^2}{m_2}}\; \cos\varphi.
\label{traj}
\end{equation}
We have to underline that the trajectory 
found using eq. (\ref{effact}) deals nothing with the
{\em real} trajectory of the second particle in the rest RF of
the first one, it expresses the mathematical solution of the
problem, defined by Lagrangian (\ref{lag}) and acquires a physical
meaning only after the proper transformation into an inertial RF.

So, we have to go back to the c.m.s.
We can note that formally we should substitute
the mass $m_2$ by the reduced mass $\mu$ in the solution.
But we can consider this problem from another side. The solution, that
we obtained, works as for {\em bound states}
as well as for {\em scattering states}. Let us look at the final
scattering state. Here we
have two free particles and, so, it is simple to go from the
rest RF of the first particle into the c.m.s. or into any RF
defined by relation (\ref{class}) using simple rules.
Here our non--inertial (in general)
reference frame appears to be
an inertial one. Let us consider the relations between
the parameters of the found solution
in our RF and in the c.m.s. A simple Galilean transformation
gives us
\begin{eqnarray} \label{rules}
E_{CM}= E_2 \frac{\mu}{m_2}\; , \qquad M_{CM}=M_2\frac{\mu}{m_2}\; ,
\qquad m_{CM}=\frac{M_{CM}^2}{2E_{CM}\rho^2}
=m_2\frac{\mu}{m_2}=\mu,
\end{eqnarray}
where $E_2=m_2v_2^2/2$ and $M_2=mv_2\rho$ are the energy and the angular momentum
of the second particle in the rest RF of the first one (in the asymptotic sate)
(they are equal to the corresponding
parameters of the whole system in the chosen RF).
The impact parameter $\rho$ is 
defined in the asymptotic initial (or final) state, it is the same
in both systems. 
$E_{CM}$ and $M_{CM}$ are the energy and the angular momentum of
the two-particle system in the c.m.s. 
Note that $m_2$ and $\mu$ in both cases
are not physical masses but some parameters of the solutions.
The energy and the angular momentum are conserved in
our problem. And we can use the relations for them as for
a scattering case as well as for a bound state, because
they {\em should} obey the
same laws being only different cases of the general solution. 
Here we rely upon analytical properties of our solution.

The substitutions defined by Eq.~(\ref{rules}) transform
the effective trajectories~(\ref{traj}) obtained from variation of
action~(\ref{effact}) into the well known trajectories
of the effective particle with the reduced mass $\mu$ in the c.m.s.
We can see that the suggested approach works in the classical
mechanics and it does not cause any difficulty.
We constructed a simple method of the two--particle
problem reduction. Now we will consider some
examples from quantum mechanics.

\section{Two--particle problem in quantum mechanics}

\subsection{Two--particle Schr\"{o}dinger equation}

Let us take the non--relativistic Schr\"{o}dinger
equation for two charged particles
\begin{eqnarray}
&& \biggl( - \vec{\nabla}_1^2 \frac{1}{2m_1}
- \vec{\nabla}_2^2 \frac{1}{2m_2} -
\frac{\alpha}{r} \biggr)\Psi(\vec{r}_1,\vec{r}_2)
= E \Psi(\vec{r}_1,\vec{r}_2),               
\label{schr}
\\  \nonumber
&& \qquad \vec{\nabla}_i^2 \equiv \frac{\partial^2}{\partial x_i^2}
+ \frac{\partial^2}{\partial y_i^2}
+ \frac{\partial^2}{\partial z_i^2}\; ,
\qquad \alpha \equiv \frac{e_1 e_2}{4 \pi }\; ,
\end{eqnarray}
where the units $c=\hbar=1$ are chosen; $e_1$ and $e_2$ are the
absolute values of the particle charges
(we assumed the attractive interaction, in the repulsion case
one has to change the sign of $\alpha$).

The equation expresses also the least action principle, it is
written just in the Hamilton form. So, as we
learned, we can try to solve it in any reference frame.
Consider again the rest reference frame of the
first particle $(\vec{p}_1=0$, $\vec{r}_1=0)$. There we have
\begin{eqnarray}
&& \Psi(\vec{r}_1,\vec{r}_2)=\Psi_1(\vec{r}_1) \Psi_2(\vec{r}_2),
\qquad \Psi_1(\vec{r}_1)=\mbox{exp}\{-i \vec{p}_1 \vec{r}_1\},
\\ && \vec{r}=\vec{r}_2-\vec{r}_1=\vec{r}_2, \qquad
E=E_2. \nonumber
\end{eqnarray}
Thus we obtain an equation of the exactly one--particle form
\begin{eqnarray}
\biggl( - \vec{\nabla}_2^2 \frac{1}{2m_2} -
\frac{\alpha}{|\vec{r}_2|} \biggr)\Psi_2(\vec{r}_2)
= E_2 \Psi_2(\vec{r}_2).               \label{schr2}
\end{eqnarray}
We should underline once more that this equation describes a non--physical
object, it has the same sense as an equation for a particle
with a {\em reduced mass}.

The discrete spectrum corresponding to eq.~(\ref{schr2}) reads
\begin{eqnarray}
E_2=-\frac{\alpha^2}{2n^2}m_2.
\end{eqnarray}
To obtain the observable spectrum we should return into
the c.m.s. We refer again to a scattering state, make
there the Galilean transformations~(\ref{rules}) and obtain just
\begin{eqnarray}
E_{CM}=-\frac{\alpha^2}{2n^2}\mu.
\end{eqnarray}

The direct relativization of the problem, i.e. the relativistic
problem for two charged scalar particles was considered in our
approach in Ref.~\cite{Arbuzov:1993qc}. There we made another
choice of a reference frame from the class given by Eq.~(\ref{class})
--- we took the RF of {\em equal velocities}
$(A=1/m_1,\ B=-1/m_2)$. The problem was explicitly reduced
to the one--particle Klein--Fock--Gordon equation. 
The comparison with the case of the rest RF choice was presented also.
The c.m.s. spectrum of this system received earlier in another approach~\cite{Tod}
was reproduced.

\subsection{One scalar and one spinor charged particles}

Let us consider now the system consisting of a scalar and
a spinor charged particles with masses $m_1$ and $m_2$,
respectively. Here we can choose either the rest frame of the 
scalar particle, or the rest frame of the spinor one. 
We should obtain coinciding observable quantities after
transition to the c.m.s. That will be a nontrivial test of our method.

We are not going to write any relativistic two--particle
equation. We are going to solve a relativistic two--particle
problem reducing it to a one--particle problem from the beginning. 
We know from the general quantum mechanics that there {\em should be}
a relativistic Schr\"{o}dinger--like two--particle equation for our
problem. But we know also that we can consider that
equation in the rest reference frame of one of the
particles. And in the chosen reference frame, we automatically 
have just the one--particle relativistic equation describing
the motion of the second particle in the field of the first one.

Let us take first the rest RF of the scalar particle.
There we have the Dirac equation for the spinor particle
in the static Coulomb potential of the scalar one:
\begin{eqnarray}
\bigl( \vec{\alpha}\vec{p}_2+e_2 A^0 + \beta m_2 \bigr) \Psi
= E_2 \Psi,
\qquad A^0=-\frac{e_1}{4\pi |\vec{r}_2|}\; .
\label{direq}
\end{eqnarray}
The exact solution of the problem is well known. For
the discrete spectrum one has
\begin{eqnarray}
&& E_2=m_2 \Biggl[ \sqrt{1+ \biggl(\frac{\alpha}{\gamma + n_r}
\biggr)^2 } \, \Biggr]^{-1}, \\  \nonumber
&& n_r=n-k, \qquad k=j+\frac{1}{2} \; , \qquad
\gamma = \sqrt{k^2 - \alpha ^2}\; ,
\end{eqnarray}
where $n$ and $j$ are the principal and the total-angular-momentum 
quantum numbers, respectively.

In a relativistic case we can also consider a scattering state
as in Eq.~(\ref{rules}) and
derive the rules of returning to the c.m.s. For wave functions
we obtain simple rules being just analogous to the well known
transformations between the laboratory system and the center-of-mass one.
To control the energy of our system it is convenient to use the
Mandelstam invariant variable $s$,
\begin{eqnarray}
s=E_{CM}^2=(p_1+p_2)^2,  \label{sss}
\end{eqnarray}
where $p_1$ and $p_2$ are the four--momenta of our particles, which can
be defined in any reference frame. So we obtain the spectrum of our two--particle
bound state in the c.m.s.
\begin{eqnarray}
E_{CM} & = & \sqrt{s}=\biggl[ m_1^2 + m_2^2 + 2m_1E_2 \biggr]^{\frac{1}{2}} = m_1 +m_2
\nonumber \\
& - & \frac{\alpha ^2}{2n^2}\mu
- \frac{\alpha ^4}{2n^3}\mu \biggl(\frac{1}{k} - \frac{3}{4n}
+ \frac{m_1 m_2}{4n(m_1+m_2)^2} \biggr)     \label{spdir}
- \frac{\alpha ^6}{2n^3}\mu \biggl[ \frac{1}{4k^3} + \frac{3}{4nk^2}
\\ \nonumber
& - & \frac{3}{2n^2k} + \frac{5}{8n^3}
+ \frac{\mu}{(m_1+m_2)}\biggl(\frac{1}{2n^2k} - \frac{3}{8n^3} \biggr)
+ \frac{\mu^2}{(m_1+m_2)^2}\,\frac{1}{8n^3} \biggr] \; + \; ...
\end{eqnarray}

Now let us consider our problem in the rest RF of the spinor
particle. Here we have the Klein--Fock--Gordon equation for
the scalar particle moving in the static potential created by the
spinor one. We have now a small additional {\em magnetic} term
in the potential due to the spin. The equation reads
\begin{eqnarray}
&& \bigl[(p_1^{\mu} - e_1 A^{\mu})^2 - m_1^2 \bigr]\Psi = 0,
\label{kg}
\\ \nonumber
&& A^0= - \frac{e_2}{4\pi r_1}\; , \qquad
\vec{A}=-\frac{[\vec{\mu} \vec{r}_1]}{4\pi r_1^3}
= - \frac{e_2}{2m_2} \cdot \frac{[\vec{s} \vec{r}_1]}{4\pi r_1^3} \; ,
\end{eqnarray}
where $\vec{\mu}$ is the magnetic moment of the spinor particle,
$\vec{s}$ is its spin, $r_1 \equiv |\vec{r}_1|$. 

It is known that the solution of the Klein--Fock--Gordon equation
with the Coulomb potential $(A^0)$ has a divergence of the
wave function in the zero point. But the additional {\em magnetic}
term $(\vec{A})$ makes the solution to be convergent in our case.
We will consider that additional term as a
perturbation, it is possible for the calculations of the spectrum.
The solution for the spectrum of Eq.~(\ref{kg}) if we took
only the Coulomb part of the potential would be the following
\begin{eqnarray}
E_1=m_1-\frac{\alpha^2}{2n^2}m_1 - \frac{\alpha^4}{2n^3}m_1
\biggl(\frac{2}{2l+1} - \frac{3}{4n}\biggr)\; + \; ...
\label{sprest}
\end{eqnarray}
The transformation into the c.m.s. gives
\begin{eqnarray}
E_{CM}' =  m_1 +m_2 - \frac{\alpha ^2}{2n^2}\mu
- \frac{\alpha ^4}{2n^3}\mu \biggl(\frac{2}{2l+1} - \frac{3}{4n}
+ \frac{m_1 m_2}{4n(m_1+m_2)^2} \biggr) \; + \; ... \; ,
\label{spcoul}
\end{eqnarray}
where $l$ is the orbital angular momentum quantum number.
The difference between spectra (\ref{spdir}) and (\ref{spcoul})
is due to the contribution of the additional
term in the potential which is not accounted in the last formula.
Let us calculate it in the perturbation theory taking
the solution of the non--relativistic Schr\"{o}dinger equation
as the ground state. 
So, we have to calculate the average value of the operator
\begin{eqnarray}
\frac{-2e_1 (\vec{p}_1 \vec{A})}{2m_1}=\frac{\alpha(\vec{p}_1
[\vec{\mu}\vec{r}_1])}{m_1 r_1^3}
= \frac{\alpha}{2m_1 m_2 r_1^3}(\vec{l_1}\vec{s})
= \frac{\alpha}{2m_1^2r_1^3}(\vec{l}\vec{s}),
\end{eqnarray}
where we divided by $2m_1$ in order to normalize the Klein--Fock--Gordon
equation to the Schr\"{o}dinger one. We used the relation between
angular momenta in different reference frames
\begin{eqnarray}
\vec{l_1}=\frac{m_2}{m_1}\, \vec{l}_{CM}\equiv \vec{l}.
\end{eqnarray}
Using the standard technique we get
\begin{eqnarray}
\delta E &=& \langle \frac{\alpha}{2m_1^2r_1^3}(\vec{l}\vec{s})
\rangle
= \frac{\alpha}{2m_1^2} \langle (\vec{l}\vec{s}) \rangle
\langle \frac{1}{r_1^3} \rangle \nonumber \\
&=& \frac{\alpha}{2m_1^2}\cdot \frac{j(j+1)-l(l+1)-s(s+1)}{2}
\cdot \frac{2\alpha ^3 m_1^3}{n^3 l(2l+1)(l+1)} \nonumber \\
&=& \frac{\alpha^4 m_1}{2 n^3}
\left\{
\begin{array}{ll}
\frac{1}{(l+1)(2l+1)} \qquad & (\mbox{for}\ j=l+\frac{1}{2}) \\[.1cm]
- \frac{1}{l(2l+1)}          & (\mbox{for}\ j=l-\frac{1}{2})
\end{array} \right.
\end{eqnarray}
The transformation to the c.m.s. (in the lowes order) leads to the substitution
$m_1 \to \mu$ in $\delta E$. Adding the received value
to the spectrum shown in eq. (\ref{spcoul}) we get the result
coinciding with spectrum (\ref{spdir}).

From this example we can see that the dependence on the reference
frame choice does not affect the
final result for the observable quantity. Calculations do not
depend on the relation between particle masses, and relativistic
recoil effects are included exactly.

\subsection{Two spinor charged particles}

Here we will very briefly show how to calculate the
positronium spectrum using our approach. In this case it
is convenient to start from the result for two scalar
particles and then add spin--orbital and spin--spin
contributions calculated as perturbations.
For equal masses in the c.m.s. from Eq.~(\ref{spcoul})
we obtain
\begin{eqnarray}
E_{CM}  =  2m - \frac{\alpha ^2}{4n^2} m
- \frac{\alpha ^4}{2n^3} m \biggl(\frac{1}{2l+1}
- \frac{11}{32n} + \varepsilon_{lSJ} \biggr) \; + \; .... \; ,
\label{sppos}
\end{eqnarray}
where $\varepsilon_{lSJ}$ is added.

The operator corresponding to the additional term is just
the well known operator of the hyperfine structure in
the hydrogen atom. And it is known (see {\it e.g.} \cite{BS}), 
that the average value of the operator for
the case of the positronium state gives the $\varepsilon$ term
for the ortho--positronium spectrum:
\begin{eqnarray}
&& \varepsilon_{l,S=0,J}=0, \nonumber \\
&& \varepsilon_{l,S=1,J}=\frac{1}{2l(2l+1)} \left\{
\begin{array}{ll}
\frac{3l+4}{(l+1)(2l+3)} \qquad & (\mbox{for}\ J=l+1) \\[.1cm]
-\frac{1}{l(l+1)}               & (\mbox{for}\ J=l)   \\[.1cm]
-\frac{3l-1}{l(2l-1)} \qquad & (\mbox{for}\ J=l-1)
\end{array} \right.
\end{eqnarray}
The contribution due to the annihilation channel can be added as a perturbation
as well.

\section{Conclusions}

The suggested approach allows to reduce any two--particle problem
with local interactions to a one--particle one. If the
particles where different one could choose the type of the
one--particle problem which is preferred. We demonstrated that the 
approach works properly as in classical mechanics as well as in 
the relativistic quantum mechanics.
In the rest RF of one particle we have to describe the motion
of the second particle in a static external potential. And there
in a relativistic case we do not have any problem with
the retardation of interactions and with the different times of
our particles, because all coordinates of the first
particle were eliminated by construction.

The two--particle problem was treated above without quantum loop
effects which are know to appear within the quantum field theory.
Here one can proceed in two wyas. The first possibility is to
choose some basic semi-calssical approximation and compute all quantum
and relativistic effects as perturbations. Actually this approach has
been widely applied to describe atomic spectra with ultimate precision~\cite{Eid}.
The other possibility is to exploit the Bethe-Salpeter approach~\cite{BSeq}.
But in both cases the choice of a reference frame should be done.
As we have demonstrated, choosing the rest reference frame of one of the
particles might have severe advantages with respect to the traditional
c.m.s. version.


The unified approach to scattering and bound states was
used. And it leads to a description of a bound state
in which the particles have some 
eigenvalues of their momenta.  And it allows
to do that simple transformations between different reference frames. 
We should note that the solution for a wave function of a bound
state found in our approach in the first particle rest RF can not
be considered as the wave function of a really moving two--particle
system, because the velocity of the system takes imaginary values.


\vspace{.5cm}

The authors expresses sincere gratitude to prof. \fbox{E.A.~Kuraev}
for fruitful discussions.

\end{document}